\documentclass[pra,twocolumn,groupedaddress,showpacs,floatfix]{revtex4}

\usepackage[utf8]{inputenc}
\usepackage{amsfonts}
\usepackage{amssymb}
\usepackage{boxedminipage}
\usepackage{listings}
\usepackage{minitoc}
\usepackage{ifpdf}

\ifpdf
\usepackage[pdftex]{graphicx}
\else
\usepackage{graphicx}
\fi

\newcommand{\ket}[1]{\left\vert#1 \right\rangle\!}
\newcommand{\bra}[1]{\!\left\langle #1\right\vert}

\begin{document}

\title{Cooling atom-cavity systems into entangled states}
\author{J. Busch,$^1$ S. De,$^1$ S. S. Ivanov,$^{1,2}$ B. T. Torosov,$^{1,2}$ T. P. Spiller,$^1$ and A. Beige$^1$}
\address{$^1$The School of Physics and Astronomy, University of Leeds, Leeds LS2 9JT, United Kingdom \\
$^2$Department of Physics, Sofia University, James Bourchier 5 blvd, 1164 Sofia, Bulgaria}

\date{\today}

\begin{abstract}
Generating entanglement by simply cooling a system into a stationary state which is highly entangled has many advantages. Schemes based on this idea are robust against parameter fluctuations, tolerate relatively large spontaneous decay rates, and achieve high fidelities independent of their initial state. A possible implementation of this idea in atom-cavity systems has recently been proposed by Kastoryano {\em et al.} [Phys.~Rev.~Lett.~{\bf 106}, 090502 (2011)]. Here we propose an improved entanglement cooling scheme for two atoms inside an optical cavity which achieves higher fidelities for comparable single-atom cooperativity parameters $C$. For example, we predict fidelities above $90 \%$ even for $C$ as low as $20$ without requiring individual laser addressing and without having to detect photons.
\end{abstract}


\maketitle

\section{Introduction}

Current atom-cavity experiments with coupling constants $g$, cavity decay rates $\kappa$, and atomic decay rates $\Gamma$ operate in a parameter regime where the single-atom cooperativity parameter $C$, 
\begin{eqnarray} \label{C}
C \equiv {g^2 \over \kappa \Gamma} \, ,
\end{eqnarray}
is at most one or two orders of magnitude larger than one \cite{Rempe0,Walther,Chapman,Meschede,Hinds,Reichel}. However, the practical realisation of atom-cavity quantum computing schemes usually requires $C$'s above $200$ to achieve single-operation fidelities above $90 \, \%$ \cite{Pellizzari,zoller2,Beige,zheng,Pachos,you,Rempe,zoller3}. The only alternative are probabilistic quantum computing schemes. These promise fidelities above $90 \, \%$ even when $C = 10$ but rely either on the detection of single photons \cite{Lim,sean} or on the observation of macroscopic fluorescence signals \cite{Metz}. Because of being conditional, they require relatively high photon detection efficiencies and cavity mirrors with low absorption coefficients. Using currently available experimental setups to entangle atoms in optical cavities with a very high fidelity therefore requires a different approach than previously proposed in the literature.

Recently it has been pointed out by several authors \cite{Milburn,Kraus-Cirac,Kraus:2008p5470,Verstraete:2009p3815,Diehl:2008p7796,Vacanti:2009p5419,Wang,Viola,Cho} that it is possible to generate entanglement in a controlled way by simply {\em cooling} qubits into well-defined, highly entangled states. The main idea behind this approach is to design laser fields such that the target state becomes the stationary state of the system. State preparation schemes based on this idea are expected to tolerate much higher spontaneous decay rates than proposals which do not use dissipation in this way. Moreover, when cooling a system into an entangled state, the fidelity of the state preparation no longer depends on the initial state of the system which makes the entanglement generation more robust against errors. Although being very promising, this approach has only recently been studied as a tool to entangle two atoms trapped inside the same optical cavity. The only examples are Kastoryano {\em et al.} \cite{Sorensen} and Wang and Schirmer \cite{Wang2}.

In this paper we follow similar ideas as Refs.~\cite{Sorensen,Wang2} and design an entangling scheme to cool two atoms inside an optical cavity into a maximally entangled state. As proposed in Ref.~\cite{Vacanti:2009p5419}, and in close analogy to the laser sideband cooling technique of trapped ions \cite{cool}, we employ level shifts and apply laser fields such that only the target state experiences off-resonant driving. Every ground state of the system other than the target state couples resonantly and sufficiently strong to rapidly decaying excited states. Doing so, the target state becomes the stationary state of the quantum system. It is reached independently of the initial state of the system after a certain transition time. As in laser sideband cooling, the fidelity of the final state reaches one when the detuning of the target state becomes much larger than the relevant laser Rabi frequencies and decay rates. 

\begin{figure}[t]
\begin{minipage}{\columnwidth}
\begin{center}
\resizebox{\columnwidth}{!}{\rotatebox{0}{\includegraphics{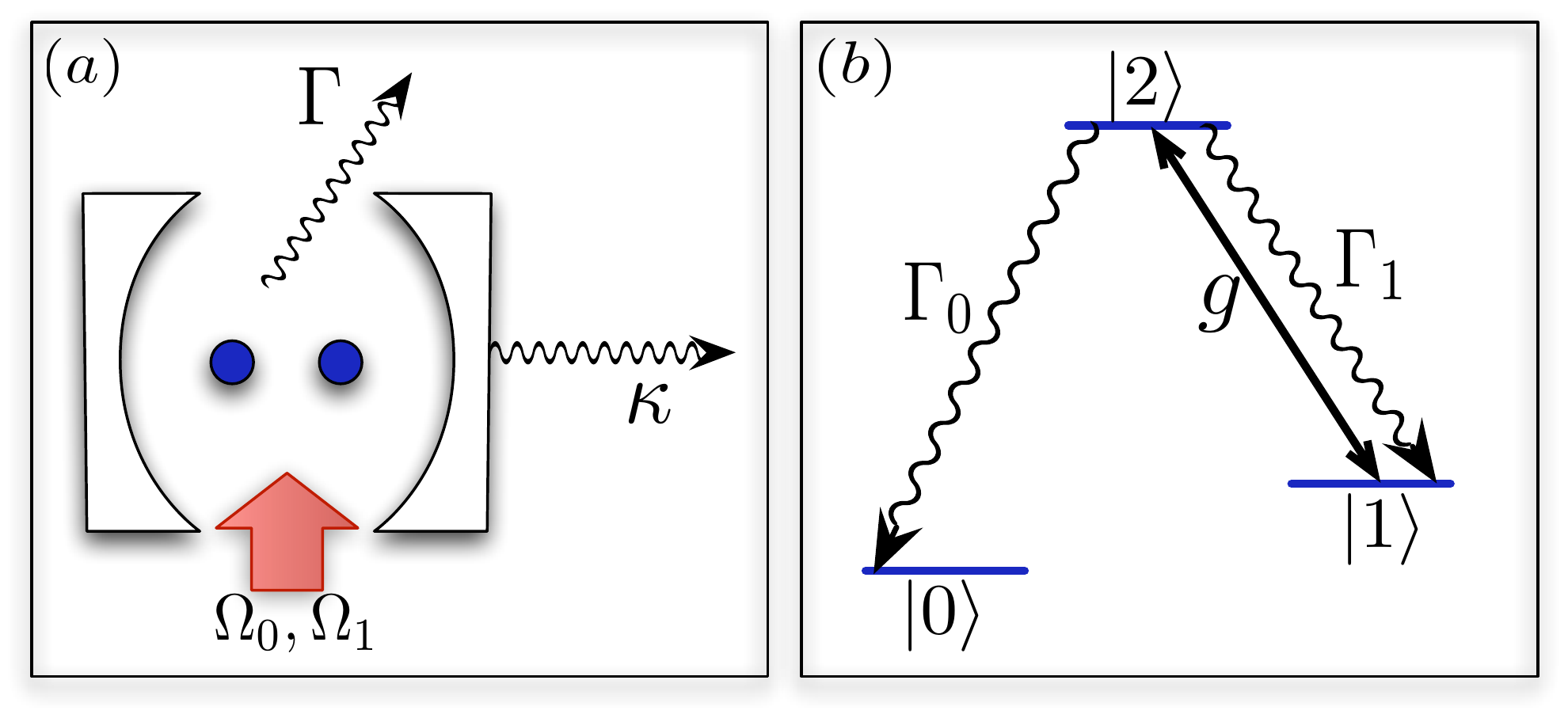}}}
\end{center}
\vspace*{-0.5cm} \caption{(a) Experimental setup to cool two atoms inside an optical cavity into a maximally entangled state. Here, $\Gamma$ and $\kappa$ denote the spontaneous atom and cavity decay rates while $\Omega_0$ and $\Omega_1$ are the relevant laser Rabi frequencies. (b) Level scheme of a single-atom. The 1--2 transition couples resonantly with coupling constant $g$ to the cavity field. The spontaneous decay rates for the 0--2 and the 1--2 transitions are $\Gamma_0$ and $\Gamma_1$ with $\Gamma = \Gamma_0 + \Gamma_1$.} \label{scheme}
\end{minipage}
\end{figure}

The concrete experimental setup which we consider in this paper is shown in Fig.~\ref{scheme}. It consists of two atoms simultaneously trapped inside an optical cavity. The main decay channels in this system are the spontaneous emission of photons from the excited atomic state $|2 \rangle$ with decay rate $\Gamma$ and the leakage of photons through the cavity mirrors with decay rate $\kappa$. Both atoms are driven by external laser fields which couple respectively to the 0--1 and the 1--2 transition. In the following, we design the detunings and Rabi frequencies of these laser fields such that the stationary state of the atom-cavity system is to a very good approximation given by the maximally entangled atomic ground state
\begin{eqnarray} \label{target}
\ket{+} &=&  (\ket{01} + \ket{10}) / \sqrt{2} \, ,
\end{eqnarray}
while there is no photon inside the cavity. As we shall see below, individual laser addressing of the atoms is not required. 
The entangling scheme proposed in this paper uses an energy shift of the target state which is due to a non-zero atom-cavity coupling constant $g$ as well as spontaneous emission from excited states. This makes it possible to prepare the state in Eq.~(\ref{target}) with a fidelity above $90 \, \%$ even when $C$ is as low as 20 and without having to detect photons.

One advantage of the state preparation scheme presented in this paper is that it predicts higher fidelities than the recently proposed entangling schemes in Refs.~\cite{Sorensen,Wang2} although they employ similar level shifts to cool two the atoms into a maximally entangled state. Ref.~\cite{Sorensen} uses a similar atomic level scheme as the one shown in Fig.~\ref{scheme} but with the addition of a driven microwave transition between the triplet states. Ref.~\cite{Wang2} relies on the presence of an external magnetic field gradient to produce the required level splittings.

There are five sections in this paper. In the next section, we introduce a four-level toy-model and discuss how to cool it into one of its ground states. The reason for the introduction of this toy-model is that the entangling scheme proposed in this paper cannot be modelled easily analytically. There is no interaction picture in which the system Hamiltonian becomes time-independent. Although being much simpler, the toy-model in Section \ref{toy_model} captures all the basic features of the proposed state preparation scheme, provides much insight into its cooling mechanism, but is nevertheless analytically tractable. In Section \ref{full}, we present all the details of our entangling scheme, draw analogies to the toy-model, and support our claims about the parameter dependence of its fidelity with the help of numerical simulations. We finally summarize our findings in Section \ref{conc}.

\section{State preparation in a toy-model} \label{toy_model}

In this section we consider a simple four-level system and pose the task to prepare it in one of its two ground states. The role of the experimental parameters in this simple model, i.e.~its laser Rabi frequencies, detunings, and spontaneous decay rates, can later be mapped onto the atom-cavity coupling constant $g$, the cavity and the atom decay rates $\kappa$ and $\Gamma$, and laser parameters $\Omega_i$ and $\delta_i$ in the entangling scheme proposed in Section \ref{full}. Our understanding of the toy-model will allow us to correctly predict the general dependence of the fidelity and the cooling rate of the proposed entangling scheme on these experimental parameters.

\subsection{Theoretical model} \label{theoretical_model}

\begin{figure}[t]
\begin{minipage}{\columnwidth}
\begin{center}
\resizebox{0.8\columnwidth}{!}{\rotatebox{0}{\includegraphics{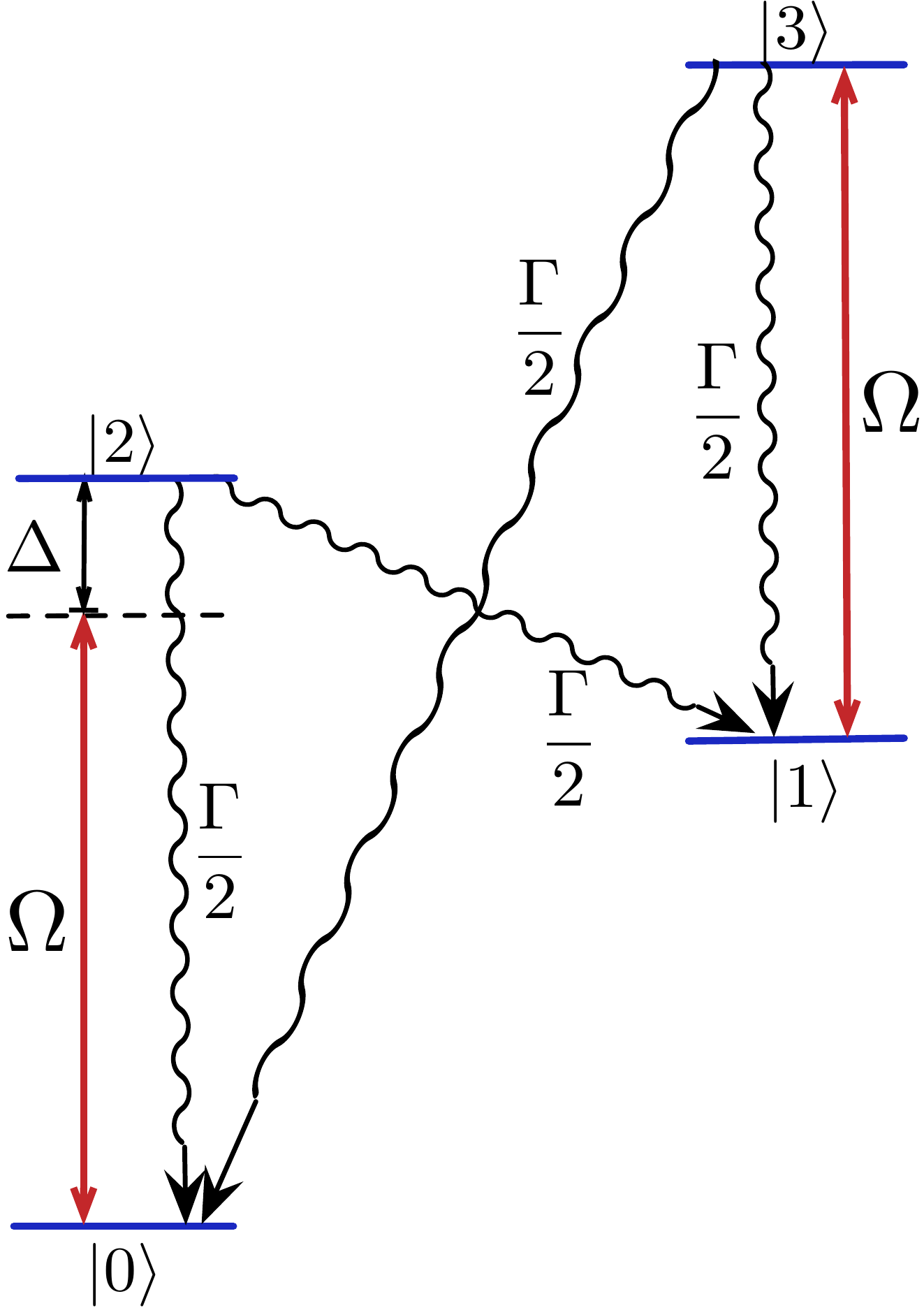}}}
\end{center}
\vspace*{-0.5cm} \caption{The toy-model level scheme. The 0--2 and the 1--3 transitions are driven by a laser field with Rabi frequency $\Omega$ and a detuning $\Delta$ with respect to the 0--2 transition. The excited atomic states both decay spontaneously into $\ket{0}$ and $\ket{1}$ with a decay rate $\Gamma/2$.} \label{toy_levels}
\end{minipage}
\end{figure}

The toy-model contains only two ground states $|0 \rangle$ and $|1 \rangle$ and two excited states $|2 \rangle$ and $|3 \rangle$, as shown in Fig.~\ref{toy_levels}. For simplicity, we assume that the decay rates for all four decay channels are the same and denote the overall spontaneous decay rate of level 2 and 3 by $\Gamma$. Moreover, we assume that the system is driven by a single laser field of frequency $\omega_{\rm L}$ and Rabi frequency $\Omega$. This laser is in resonance with the 1--3 transition but detuned from the 0--2 transition by a detuning $\Delta$. The spontaneous emission of photons is in the following taken into account using the master equation 
\begin{eqnarray}\label{t_master}
	\dot{\varrho}(t) &=& -\frac{{\rm i}}{\hbar} \left[H_{\rm cond} \varrho - \varrho H_{\rm cond}^{\dagger}\right] + \mathcal{R}(\varrho) \, ,
\end{eqnarray}
where $\varrho $ is the density matrix of the system. The conditional Hamiltonian $H_{\rm cond}$ describes the time evolution under the condition of no photon emission, while ${R}(\varrho) $,
\begin{eqnarray}
\mathcal{R}( \varrho ) &=& \sum_{i=0,1} \sum_{j=2,3} \frac{1}{2} \Gamma \ket{i}\bra{j}\varrho\ket{j}\bra{i} \, ,
\end{eqnarray} 
relates to the reset state in case of a photon emission. Within the rotating wave approximation and in the interaction picture with respect to the free Hamiltonian
\begin{eqnarray}\label{toy_H0}
	H_0 &=& \sum_{i=0}^3 \hbar \omega_i \ket{i}\bra{i} - \hbar \Delta \ket{2}\bra{2} \, ,
\end{eqnarray}
where $\hbar \omega_i$ is the energy of level $i$, $H_{\rm cond}$ equals
\begin{eqnarray}\label{toy_Hcond}
	H_{\rm cond} &=& \frac{1}{2} \hbar\Omega \left(\ket{0}\bra{2} + \ket{1}\bra{3} + {\rm H.c.} \right) + \hbar\Delta \ket{2}\bra{2} \nonumber \\
		&& - {1 \over 2} {\rm i}\hbar\Gamma \left( \ket{2}\bra{2} + \ket{3}\bra{3} \right) \, ,
\end{eqnarray}
since $\omega_{\rm L} = \omega_2 - \Delta$.  

The master equation in Eq.~(\ref{t_master}) can now be used to calculate the fidelity of the proposed state preparation scheme, i.e.~the stationary state population of its target state $|0 \rangle$, and its cooling rate. This is most easily done using rate equations which are a complete set of differential equations for the time evolution of expectation values. The time derivative of an expectation value of a time-independent operator $A$ equals 
\begin{eqnarray}
\langle \dot A \rangle &=& \mbox{Tr} (A \dot{\varrho}) \, .
\end{eqnarray}
The above master equation hence implies that
\begin{eqnarray} \label{dotA0}
\langle \dot A \rangle &=&- {1 \over 2} {\rm i} \Omega \left \langle \, \left[ A , |0 \rangle \langle 2| + |1 \rangle \langle 3| + {\rm H.c.} \right] \, \right \rangle \nonumber \\
&& - {\rm i} \Delta \left \langle \, \left[ A , |2 \rangle \langle 2| \, \right] \, \right \rangle - \sum_{j=2,3} {1 \over 2} \Gamma \left \langle \, A |j \rangle \langle j| + |j \rangle \langle j| A \, \right \rangle \nonumber \\
&& + \sum_{i=0,1} \sum_{j=2,3} {1 \over 2} \Gamma \left \langle \, |j \rangle \langle i| \, A \, |i \rangle \langle j| \, \right \rangle \, . ~~
\end{eqnarray}
In the following we consider the Hermitian operators $|i \rangle \langle i|$, $|i \rangle \langle j| + |j \rangle \langle i|$, and ${\rm i} (|i \rangle \langle j| - |j \rangle \langle i| )$ and denote their (real) expectation values by
\begin{eqnarray}
	P_i &=& \bra{i}\varrho\ket{i} \, , \nonumber \\
	k_{ij} &=& 2{\rm Im} \, \bra{i}\varrho\ket{j} \, , \nonumber \\
	l_{ij} &=& 2{\rm Re} \, \bra{i}\varrho\ket{j} \, .
\end{eqnarray}
Substituting these operators into Eq.~(\ref{dotA0}) yields 
\begin{eqnarray} \label{rateeqns}
	\dot{P}_0 &=& - {1 \over 2} \Omega \, k_{02} + {1 \over 2} \Gamma \left( P_2 + P_3 \right) \, , \nonumber \\
	\dot{P}_1 &=&  - {1 \over 2} \Omega \, k_{13} + {1 \over 2} \Gamma \left( P_2 + P_3 \right) \, , \nonumber \\
	\dot{P}_2 &=& {1 \over 2} \Omega \, k_{02} - \Gamma P_2 \, , \nonumber \\
	\dot{P}_3 &=& {1 \over 2} \Omega \, k_{13} - \Gamma P_3 \, , \nonumber \\
	\dot{k}_{02} &=& \Omega (P_0 - P_2) + \Delta l_{02} - {1 \over 2} \Gamma k_{02} \, , \nonumber \\
	\dot{k}_{13} &=& \Omega (P_1 - P_3) - {1 \over2} \Gamma k_{13} \, , \nonumber \\ 
	\dot{l}_{02} &=& -\Delta k_{02} - {1 \over 2} \Gamma l_{02} \, .
\end{eqnarray}
These seven equations form a complete set of rate equations and are sufficient to analyse the time evolution of the population $P_0$ in the target state $|0 \rangle$ which equals the fidelity ${\rm F}$.   

\subsection{The basic idea} \label{basic_idea}

Suppose we aim to transfer the toy-model in Fig.~\ref{toy_levels} into one of its ground states, for example the $|0 \rangle$ state, with a very high fidelity and without having to control the initial state of the system. This is possible when laser driving is applied such that the $|0 \rangle$ state becomes the stationary state of the toy-model as it applies when the laser detuning for the $0$--$2$ transition is much larger than the other system parameters, i.e.~when
\begin{eqnarray}\label{toy_regime}
\Delta &\gg & \Omega, \, \Gamma \, .
\end{eqnarray}
This condition guarantees that it is much more likely for the system to spontaneously decay into the $|0 \rangle$ state when being in one of the other three states, than being driven out of it \cite{Vacanti:2009p5419}. Once the system has reached its stationary state it therefore remains there with a very high probability. 

\begin{figure}[t]
\begin{minipage}{\columnwidth}
\begin{center}
\resizebox{\columnwidth}{!}{\rotatebox{0}{\includegraphics{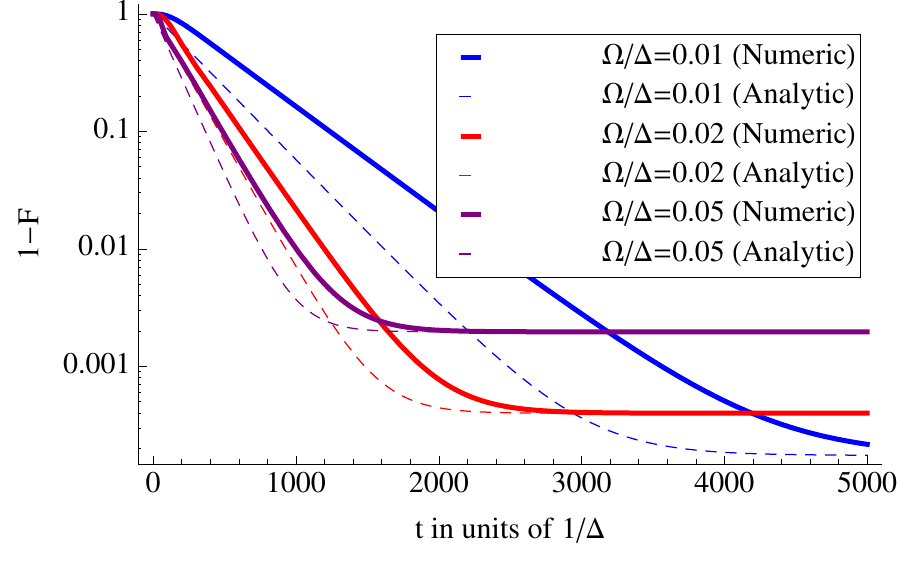}}}
\resizebox{\columnwidth}{!}{\rotatebox{0}{\includegraphics{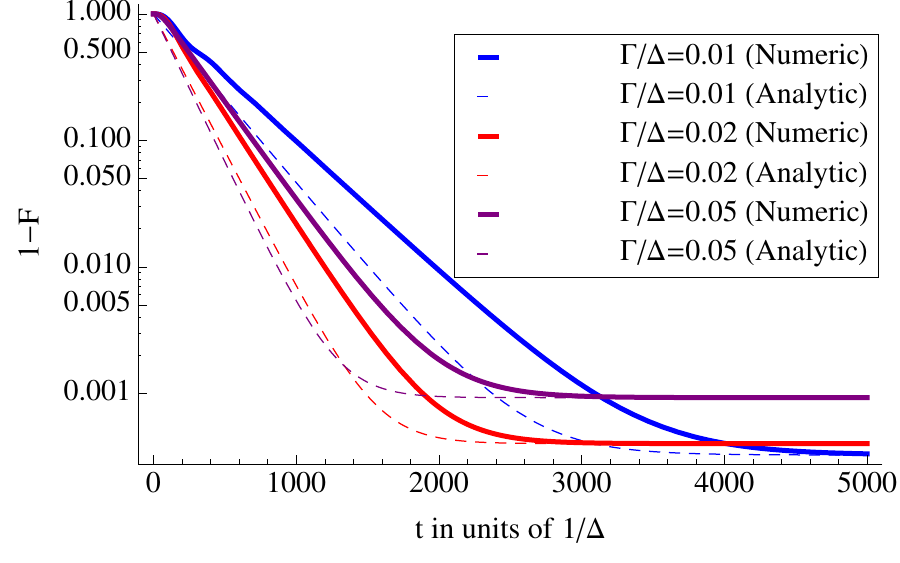}}}
\end{center}
\vspace*{-0.5cm} \caption{Logarithmic plot of the time dependence of the distance $1-{\rm F}$ from the target state $|0 \rangle$ for different values of $\Omega/\Delta$ and $\Gamma/\Delta$. The system is initially in $\ket{1}$. In the upper plot we have $\Gamma = 0.2 \, \Delta$. In the lower plot we have $\Omega = 0.05 \, \Delta$. The solid lines are the numerical solutions of the rate equations in Eq.~(\ref{rateeqns}). The dashed lines illustrate the analytical solution in Eq.~(\ref{F(t)}).} \label{toy_omega}
\end{minipage}
\end{figure}

This is confirmed by Fig.~\ref{toy_omega} which shows the time dependence of $1-{\rm F}$, i.e.~of the total population in states other than the target state $|0 \rangle$, for a wide range of experimental parameters. The solid lines in this Fig.~\ref{toy_omega} are the result of a numerical integration of the rate equations in Eq.~(\ref{rateeqns}) which assumes the worst case scenario with the toy-model initially in $|1 \rangle$. The plots show exponential cooling towards the target state until the system reaches a stationary state. The fidelity of the state preparation equals the population of the $|0 \rangle$ state and is indeed very close to unity, as long as condition (\ref{toy_regime}) applies. The cooling rate and the fidelity of the state preparation both depend on the relative size of $\Omega$ and $\Gamma$ with respect to the detuning $\Delta$. 

\subsection{Stationary state fidelity} \label{toy_fidelity}

\begin{figure}[t]
\begin{minipage}{\columnwidth}
\begin{center}
\resizebox{\columnwidth}{!}{\rotatebox{0}{\includegraphics{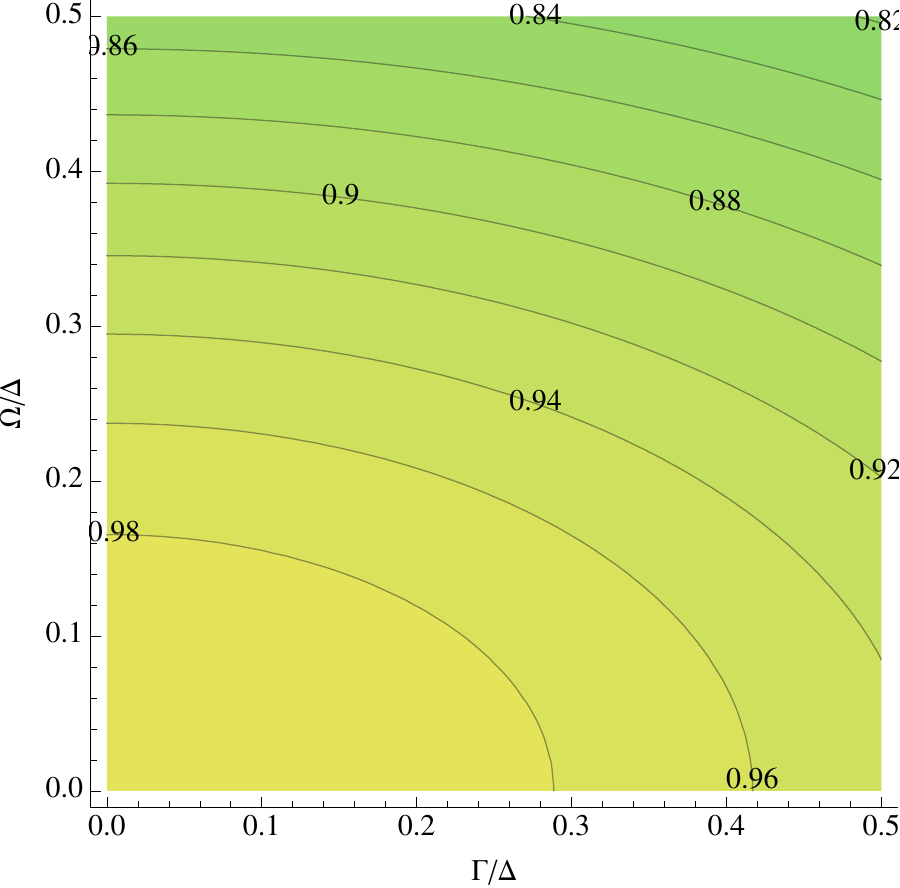}}}
\end{center}
\vspace*{-0.5cm} \caption{Contour plot which shows the stationary state fidelity ${\rm F}$ in Eq.~(\ref{t_stationary}) as a function of $\Omega /\Delta $ and $\Gamma /\Delta$.} \label{toy_contour}
\end{minipage}
\end{figure}

To identify the best way of preparing the target state, we now derive approximate analytical expressions for the stationary state fidelity, ${\rm F}$, and the cooling rate, $\gamma_{\rm c}$, which is a measure for the time it takes the system to reach its stationary state. Since  ${\rm F}$ is the stationary state population $P_0$ in the $\ket{0}$ state, it can be calculated simply by setting the time derivatives of the expectation values in Eq.~(\ref{rateeqns}) equal to zero. Doing so, we find that
\begin{eqnarray}\label{t_stationary}
	{\rm F} &=& 1 - \frac{3\Omega^2 + \Gamma^2}{4\Delta^2 + 4\Omega^2 + 2\Gamma^2} \, .
\end{eqnarray}
For relatively large detunings $\Delta$, as in Eq.~(\ref{toy_regime}), this equation simplifies to
\begin{eqnarray}\label{t_stationary2}
	{\rm F} &=& 1 - \frac{3\Omega^2 + \Gamma^2}{4\Delta^2} \, .
\end{eqnarray}
This result confirms that the fidelity is close to one in the parameter regime given by Eq.~(\ref{toy_regime}). Fig.~\ref{toy_contour} illustrates the effects of finite $\Omega$ and $\Gamma$. It also shows that an increase in the Rabi frequency $\Omega$ reduces the stationary state fidelity more rapidly than an increase in the decay rate $\Gamma$. 

\subsection{Heating and cooling rates} \label{cooling_rate}   

To see how quickly the toy-model reaches its stationary state, we now introduce the notion of a cooling and a heating rate which we denote $\gamma_{\rm c}$ and $\gamma_{\rm h}$, respectively. For simplicity, and since we are anyway only interested in the general scaling of these rates with the experimental parameters, we assume that these rates do not depend on the current state $\varrho $ of the system. Invoking the conservation of probability flux, we then find that 
\begin{eqnarray}\label{gammas}
	\dot{P}_0 = \gamma_{\rm c} \, \left( 1-P_0 \right) - \gamma_{\rm h} \, P_0 \, .
\end{eqnarray}
The principle here is that the rate at which the fidelity, i.e.~the current population in the $|0 \rangle$ state, changes in time is equal to the rate at which population is cooled into the target state minus the rate at which population is heated out of the target state. When the system reaches its stationary state, the fidelity remains constant. The above equation hence implies
\begin{eqnarray}\label{gamma_stationary}
	\gamma_{\rm h} \, {\rm F} = \gamma_{\rm c} \, \left( 1 - {\rm F} \right) \, .
\end{eqnarray}
Since we already know ${\rm F}$ (cf.~Eq.~(\ref{t_stationary})), this relation can be used to obtain the cooling rate after obtaining an estimate for the heating rate. As we shall see below, it is easier to derive an approximate expression for  $\gamma_{\rm h}$, than calculating $\gamma_{\rm c}$ directly.

\begin{figure}[t]
\begin{minipage}{\columnwidth}
\begin{center}
\resizebox{\columnwidth}{!}{\rotatebox{0}{\includegraphics{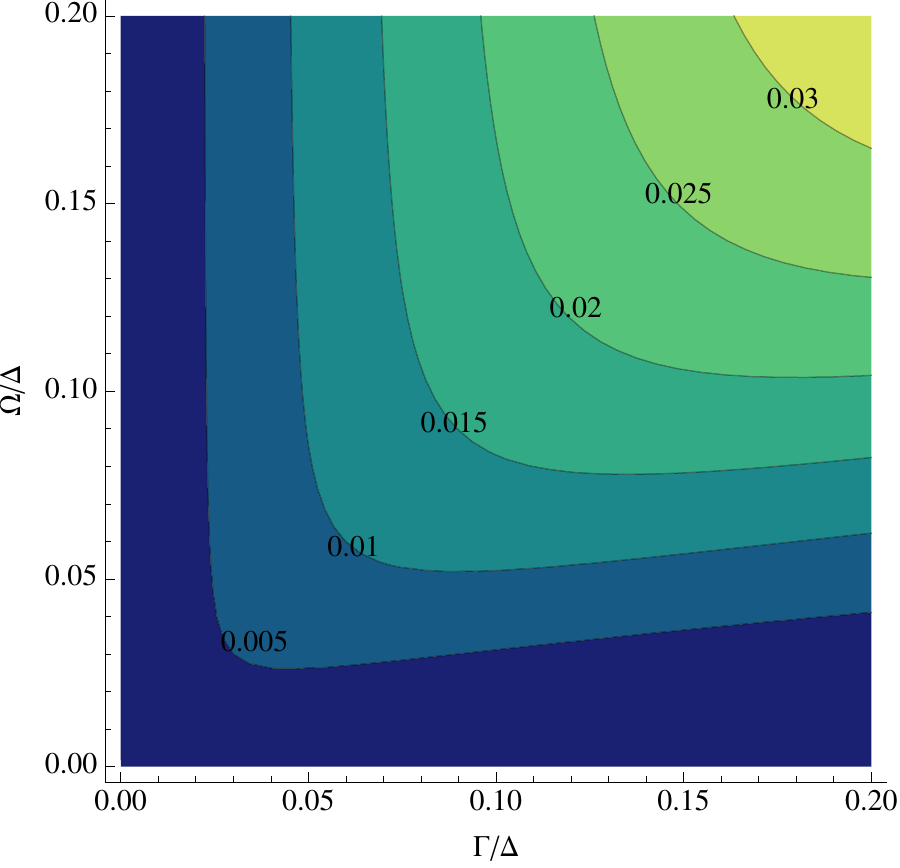}}}
\end{center}
\vspace*{-0.5cm} \caption{Contour plot which shows the cooling rate $\gamma_{\rm c}$ in Eq.~(\ref{gamma_c}) as a function of $\Omega / \Delta$ and $\Gamma / \Delta$.} \label{toy_CoolRate}
\end{minipage}
\end{figure}

Considering the parameter regime in Eq.~(\ref{toy_regime}), the rate equations in Eq.~(\ref{rateeqns}) can be simplified via an adiabatic elimination. Only the coherences ${k}_{02}$ and ${l}_{02}$ evolve on the fast time scale given by $\Delta$. Setting their time derivatives equal to zero, we find that
\begin{eqnarray}
	k_{02} = \frac{2 \Gamma\Omega}{\Gamma^2 + 4 \Delta^2}\left( P_0 - P_2 \right) \, .
\end{eqnarray}
Assuming that the toy-model is in $|0 \rangle$, i.e.~that $P_0 = 1$ and $P_1 = P_2 = P_3 = 0$, and substituting the above expression for $k_{02}$ into the rate equation for $P_0$ yields
\begin{eqnarray}\label{toy_P_0}
	\dot{P}_0 = - \frac{\Gamma\Omega^2}{\Gamma^2 + 4\Delta^2}\, P_0 \, .
\end{eqnarray}
Comparing this equation with Eq.~(\ref{gammas}) for $P_0 =1$, we find that the heating rate is to a very good approximation given by
\begin{eqnarray}\label{gamma_h}
	\gamma_{\rm h} = \frac{\Gamma\Omega^2}{\Gamma^2 + 4\Delta^2} \, .
\end{eqnarray}
This is the rate at which the target state $|0 \rangle$ loses its population. Substituting this result and Eq.~(\ref{t_stationary}) into Eq.~(\ref{gamma_stationary}), we get
\begin{eqnarray} \label{gamma_c}
	\gamma_{\rm c} &=& \frac{\Gamma \Omega^2 \left( 4\Delta^2 + \Omega^2 + \Gamma^2 \right)}{\left( 4\Delta^2 + \Gamma^2 \right)\left( 3\Omega^2 + \Gamma^2 \right)} \, .
\end{eqnarray}
Fig.~\ref{toy_CoolRate} shows this cooling rate $\gamma_{\rm c}$ for a wide range of experimental parameters. For relatively small Rabi frequencies $\Omega$, the cooling process becomes faster with increasing $\Omega$. However, it is not worth increasing $\Omega$ beyond a certain size which saturates $\gamma_{\rm c}$. In the parameter regime given by Eq.~(\ref{toy_regime}), the cooling rate $\gamma_{\rm c}$ simplifies to
\begin{eqnarray} \label{gamma_c2}
	\gamma_{\rm c} &=& \frac{\Gamma \Omega^2}{3\Omega^2+\Gamma^2} \, .
\end{eqnarray}
which no longer depends on $\Delta$ but only holds for sufficiently large detunings. 

In order to get a feeling for the accuracy of the cooling rate $\gamma_{\rm c}$ in Eq.~(\ref{gamma_c}), we now solve Eq.~(\ref{gammas}) analytically and compare the result with exact numerical solutions of the rate equations in Eq.~(\ref{rateeqns}). Doing so and assuming $P_0(0) = 0$ we find that 
\begin{eqnarray}\label{F(t)}
	P_0(t) = \frac{\gamma_{\rm c}}{\gamma_{\rm c}+\gamma_{\rm h}}\left( 1 - {\rm e}^{-\left(\gamma_{\rm c} + \gamma_{\rm h}\right)t} \right) \, .
\end{eqnarray}
Fig.~\ref{toy_omega} compares this analytical result with numerical solutions of $P_0(t)$ for different experimental parameters $\Omega/\Delta$ and $\Gamma/\Delta$. It shows that Eq.~(\ref{gamma_c}) reflects the general parameter dependence of the cooling rate on $\Omega/\Delta$ and $\Gamma/\Delta$ correctly. The above approximate solution is in general slightly higher than the actual cooling rate. The reason for this is that the heating rate in Eq.~(\ref{gamma_h}) has been calculated for the case, where the system is initially in $|0 \rangle$, i.e.~when it is the highest.

\subsection{Choosing experimental parameters} \label{choosing}

Fig.~\ref{toy_contour} shows that maximising the stationary state fidelity ${\rm F}$ requires a Rabi frequency $\Omega$ as small as possible. However, from Fig.~\ref{toy_CoolRate} we see that we only obtain high cooling rates when $\Omega$ is relatively large. To minimise the state preparation time while maintaining a high fidelity, we therefore suggest using a laser pulse with a time-dependent Rabi frequency to prepare the target state. This laser pulse should be large initially and should reach zero by the end of the cooling process. For example one could choose
\begin{eqnarray}\label{toy_Ot}
	\Omega(t) = {3 \Omega_0 \over \left( 1 + \gamma_{\rm c}(0) t \right)^{2}} 
\end{eqnarray}
with $\gamma_{\rm c}(0)$ being the cooling rate in Eq.~(\ref{gamma_c2}) for the initial Rabi frequency $\Omega(0) = \Omega_0$. Alternatively, one could choose an exponentially decreasing Rabi frequency. However, in this case, $\Omega$ would drop off too rapidly, thereby resulting in a fidelity that is far from optimal. Here we do not discuss how to optimise the spontaneous decay rate, since $\Gamma$ is in general fixed.

\begin{figure}[t]
\begin{minipage}{\columnwidth}
\begin{center}
\resizebox{\columnwidth}{!}{\rotatebox{0}{\includegraphics{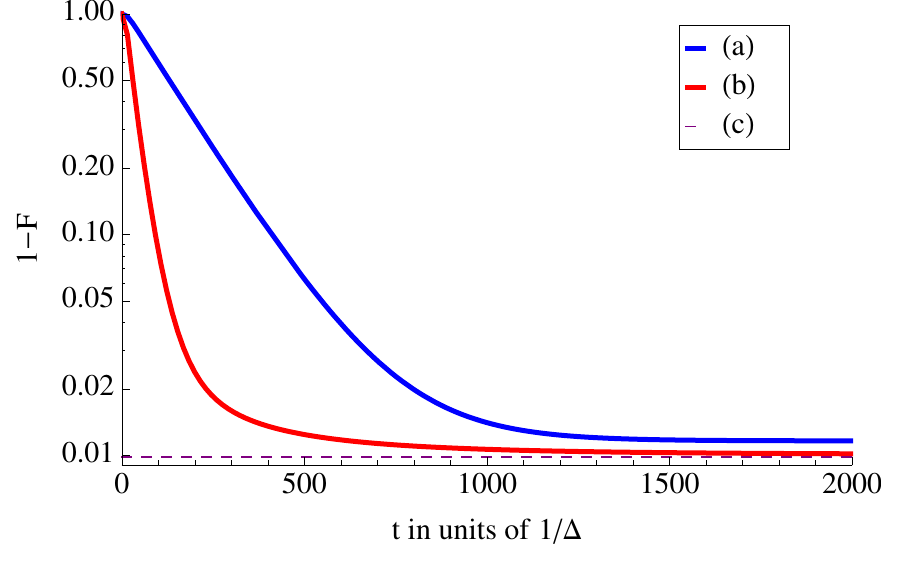}}}
\end{center}
\vspace*{-0.5cm} \caption{(a) Logarithmic plot of the time dependence of the distance $1-{\rm F}$ from the target state $|0 \rangle$ for the case where the system is initially in $\ket{1}$ and where $\Gamma = 0.2 \, \Delta$ and $\Omega = 0.05 \, \Delta$. (b) Same as (a) but for a time dependent laser pulse with a Rabi frequency $\Omega(t)$ as in Eq.~(\ref{toy_Ot}) with $\Omega_0 = 0.05 \, \Delta$. (c) Theoretical minimum for $1-{\rm F}$ obtained from Eq.~(\ref{opt}) for $\Gamma = 0.2 \, \Delta$.} \label{toy_varOmega}
\end{minipage}
\end{figure}

Fig.~\ref{toy_varOmega} confirms that choosing the Rabi frequency $\Omega$ as in Eq.~(\ref{toy_Ot}) indeed yields a significant speed up compared to time-independent Rabi frequencies. We also observe a stationary state fidelity which is close to the theoretical maximum obtained when setting $\Omega = 0$. From Eq.~(\ref{t_stationary2}) we see that this maximum is to a very good approximation given by
\begin{eqnarray} \label{opt}
	{\rm F}(\Omega=0) = 1 - \frac{\Gamma^2}{4 \Delta^2 + 2 \Gamma^2} \, .
\end{eqnarray}
It is indicated by a dashed line in Fig.~\ref{toy_varOmega}. In the next Section we use similar time-dependent laser pulses to prepare two atoms inside an optical cavity relatively fast and with a high fidelity in a maximally entangled state.

\section{Entangling Scheme} \label{full}

The toy-model described in the previous section is based on a simple principle: driving populations out of all undesired states resonantly while driving the target state off-resonantly. This approach can indeed be used to prepare target states with a high fidelity \cite{Vacanti:2009p5419}. In this section, we use this idea to prepare two atoms inside an optical cavity (cf.~Fig.~\ref{scheme}) in the maximally entangled state $|+ \rangle$ in Eq.~(\ref{target}). The first half of this section presents a theoretical description of the atom-cavity system. After identifying its dressed states, we select appropriate laser Rabi frequencies and detunings. As we shall see below, the state preparation requires three different driving lasers but there is no need to address atoms individually. A comparison with the toy-model introduced in the previous section allows us to predict the dependence of fidelity and cooling rate of the proposed state preparation scheme on the experimental parameters with a very high accuracy. 

\subsection{System Hamiltonian without laser driving} \label{AtomCav_theoretical_model}

The experimental setup which we consider in this paper consists of two atoms placed inside an optical cavity as shown in Fig.~\ref{scheme}. Each atom contains a $\Lambda$-type level configuration with $\hbar \omega_j$ and $|j \rangle$  denoting the corresponding energies and energy eigenstates $(j=0,1,2)$. Suppose the 1--2 transition of each atom couples resonantly with coupling strength $g$ to the quantised cavity field mode with frequency $\omega_{\rm c}$. Then the Hamiltonian $H_{\rm sys}$ of this system equals 
\begin{eqnarray}\label{Hsys}
H_{\rm sys} &=& \sum_{i=1}^{2}\hbar g \, |1\rangle_{ii}\langle2| c^{\dagger}+{\rm H.c.} + \sum_{i=1}^{2} \sum_{j=0}^2 \hbar\omega_{j} \, |j\rangle_{ii}\langle j\vert \nonumber \\
&& + \hbar \omega_{\rm c} \, c^{\dagger}c 
\end{eqnarray}
in the absence of external laser driving. Here $c$ and $c^\dagger $ are the cavity photon annihilation and creation operators for a single photon inside the optical cavity. As we shall see below, it is important that the atomic states $|0 \rangle$ and $|1 \rangle$ differ in energy by an amount which is significantly larger than $\hbar g$. 

\begin{table}[t]
\begin{center}
\begin{tabular}{| c | c |}
	\hline 
	~~ Energy eigenstate ~~ & ~~ Energy ~~ \\ \hline \hline
	$\ket{00,0}$ & 0 \\
	~~ $\ket{+,0} \equiv \left(\ket{01,0} + \ket{10,0}\right) / \sqrt{2} $ ~~ & $\hbar \omega_1$ \\
	~~ $\ket{-,0} \equiv \left(\ket{01,0} - \ket{10,0}\right) / \sqrt{2} $ ~~ & $\hbar \omega_1$ \\
	$\ket{11,0}$ & ~~ $2 \hbar \omega_1$ ~~ \\
	\hline
\end{tabular}
\end{center}
\caption{Energy eigenstates and energy eigenvalues of the system Hamiltonian $H_{\rm sys}$ in Eq.~(\ref{Hsys}) for the ${\cal H}_0$ subspace with no atom in $|2 \rangle$ and no photons in the cavity.} \label{g_states}
\end{table}

In the following, we identify the relevant energy eigenstates of this Hamiltonian, since this will allow us to identify appropriate laser drivings and detunings for the proposed state preparation scheme. To do so, we denote states with atom $1$ in $\ket{j_1}$, atom $2$ in $\ket{j_2}$ and $n$ photons in the cavity by $\ket{j_1j_2,n}$. Moreover we notice that the Hamiltonian $H_{\rm sys}$ preserves the total amount of population in the excited atomic state $|2 \rangle$ and the cavity field mode. It therefore acts on fixed excitation subspaces ${\cal H}_n$ of the complete Hilbert space whose energy eigenstates can be calculated separately. The eigenstates and eigenvalues of the subspace ${\cal H}_0$ of states with no population of the excited atomic state $|2 \rangle$ and no photons in the cavity are summarised in Table~\ref{g_states}. Table~\ref{e_states} shows the eight energy eigenstates and the corresponding energy eigenvalues of the subspace ${\cal H}_1$ of states with either one atom in $|2 \rangle$ or one photon in the cavity and adopts the notation
\begin{eqnarray}\label{eigenvecs}
	\ket{\mu_{1}} &\equiv & \left(\ket{21,0} - \ket{12,0}\right) / \sqrt{2} \, , \nonumber \\
	\ket{\mu_{0},\pm} &\equiv & \left(\ket{02,0} - \ket{20,0} \pm \ket{01,1} \mp \ket{10,1}\right) / 2 \, , \nonumber \\
	\ket{\lambda_{0},\pm} &\equiv & \left(\ket{02,0} + \ket{20,0} \pm \ket{01,1} \pm \ket{10,1}\right) / 2 \, , \nonumber \\
	\ket{\lambda_{1},\pm} &\equiv & \left(\ket{12,0} + \ket{21,0} \pm \sqrt{2}\ket{11,1}\right) / 2 \, .
\end{eqnarray}
Fortunately, there is no need to identify the energy eigenstates of the atom-cavity system of the subspace of states with more than one excitation in the atomic state $|2 \rangle$ and the cavity field mode. The reason for this is that these states do not couple directly to the states in Table~\ref{g_states} in case of laser driving. Therefore they do not have to be taken into account when choosing laser parameters such that only the $|+,0 \rangle$ state experiences off-resonant laser driving.

\subsection{Laser driving} \label{laserdriving}

As we shall see below, the state preparation of the maximally entangled atomic state $|+ \rangle$ in Eq.~(\ref{target}) requires the simultaneous excitation of the two atoms with three different laser fields. In the following we assume that the 0--2 transition of each atom is driven by two different lasers with Rabi frequencies $\Omega_{0}^{(k)}$ and frequencies $\omega_{0}^{(k)}$ respectively $(k=1,2)$. The 1--2 transition of each atom should moreover be driven by a laser field with Rabi frequency $\Omega_1$ and frequency $\omega_{\rm L1}$. The laser Hamiltonian in the Schr\"odinger picture and the usual rotating wave approximation is then given by
\begin{eqnarray} \label{HL}
	H_{\rm L}(t) &=& \sum_{i=1}^{2} \sum_{k=1}^{2} \frac{1}{2} \hbar \Omega^{(k)}_{0} \, {\rm e}^{{\rm i} \omega^{(k)}_{\rm L0} t} \, |0 \rangle_{ii} \langle 2| + {\rm H.c.} \nonumber \\
	 && + \sum_{i=1}^{2} \frac{1}{2} \hbar \Omega_{1} \, {\rm e}^{{\rm i} \omega_{\rm L1} t} \, |1 \rangle_{ii} \langle 2| + {\rm H.c.}   
\end{eqnarray}
The realisation of this Hamiltonian does not require individual laser addressing, since both atoms experience exactly the same laser driving. 

\begin{table}[t]
\begin{tabular}{| c | c |}
	\hline ~~ Energy eigenstate ~~ & ~~ Energy ~~ \\ \hline \hline
	$\ket{00,1}$ & ~~ $\hbar \omega_{\rm c} = \hbar (\omega_2 - \omega_1)$ ~~ \\
	$\ket{\mu_{1}}$ & $\hbar (\omega_1 + \omega_2)$ \\
	$\ket{\mu_{0},\pm}$ & $\hbar (\omega_2 \pm g)$ \\
	$\ket{\lambda_{0},\pm}$ & $\hbar (\omega_2 \pm g)$ \\
	$\ket{\lambda_{1},\pm}$ & $\hbar (\omega_1 + \omega_2 \pm \sqrt{2}g)$ \\
	\hline
\end{tabular}
\caption{Energy eigenstates and energy eigenvalues of the system Hamiltonian $H_{\rm sys}$ in Eq.~(\ref{Hsys}) for the ${\cal H}_1$ subspace of states with either one atom in $|2 \rangle$ or one photon in the cavity. The table uses the notation introduced in Eq.~(\ref{eigenvecs}).} \label{e_states}
\end{table}

In order to see how to best choose the laser frequencies $\omega_{\rm L0}^{(k)}$ and $\omega_{\rm L1}$, we now consider the effect of this laser Hamiltonian on the ${\cal H}_0$ subspace. This effect can be described by the restricted laser Hamiltonian $\tilde H_{\rm L}$ defined as 
\begin{eqnarray}
	\tilde H_{\rm L}(t) &\equiv & P \, H_{\rm L}(t) \, P 
\end{eqnarray}
with the projector $P$ being the projector on ${\cal H}_0$ and ${\cal H}_1$ given by 
\begin{eqnarray}
P &=& \sum_{x=+,-}  |\mu_0,x \rangle \langle \mu_0,x| + \sum_{j=0,1} \sum_{x=+,-} |\lambda_j,x \rangle \langle \lambda_j,x| \nonumber \\
&& + |\mu_1 \rangle \langle \mu_1|\, .
\end{eqnarray}
Using the eigenvectors of the undriven atom-cavity system Hamiltonian which can be found in Tables~\ref{g_states} and \ref{e_states} one can show that this Hamiltonian equals
\begin{eqnarray}\label{H_laser}
	\tilde H_{\rm L}(t) &=& \sum_{k=1}^{2} \sum_{x=+,-} {1 \over 2 \sqrt{2}} \hbar \Omega_0^{(k)} \, {\rm e}^{{\rm i}\omega_{{\rm L0}}^{(k)}t} \, \Big[ \ket{00,0} \bra{\lambda_0,x} \nonumber \\
	&& + \ket{+,0} \bra{\lambda_1,x} \Big] + \sum_{k=1}^{2} {1 \over 2} \hbar \Omega_0^{(k)} \, {\rm e}^{{\rm i}\omega_{{\rm L0}}^{(k)}t} \, \ket{-,0}\bra{\mu_1} \nonumber \\
	&& + \sum_{x=+,-} \frac{1}{2 \sqrt{2}} \hbar \Omega_1 \, {\rm e}^{{\rm i}\omega_{\rm L1} t} \, \Big[ \ket{+,0} \bra{\lambda_0,x} \nonumber \\
	&& + \ket{-,0} \bra{\mu_0,x} + \ket{11,0} \bra{\lambda_1,x} \Big] + {\rm H.c.} 
\end{eqnarray}
in the Schr\"odinger picture. Changing into an interaction picture in which $\tilde H_{\rm L}(t)$ becomes time independent is not possible, since there are more laser fields than atomic transitions. 

\begin{table}[t]
	\begin{tabular}{| c | c | c | c |}
		\hline
		\hspace*{0.05cm} Ground \hspace*{0.05cm} & \hspace*{0.05cm} Excited \hspace*{0.05cm} & Rabi & \hspace*{0.05cm} Effective \hspace*{0.05cm} \\[-0.15cm]
		state & state  & \hspace*{0.05cm} frequency \hspace*{0.05cm} & \hspace*{0.05cm} detuning \hspace*{0.05cm} \\[0.1cm] \hline \hline
		$\ket{00,0}$ & $\ket{\lambda_0,\pm}$ & $\Omega_0^{(1)} / \sqrt{2}$ & \hspace*{0.05cm} $\omega_{\rm L0}^{(1)} - \omega_2 \pm g$ \hspace*{0.05cm} \\
		                    & $\ket{\lambda_0,\pm}$ & $\Omega_0^{(2)} / \sqrt{2}$ & \hspace*{0.05cm} $\omega_{\rm L0}^{(2)} - \omega_2 \pm g$ \hspace*{0.05cm} \\ \hline
		$\ket{+,0}$ & $\ket{\lambda_1,\pm}$ & $\Omega_0^{(1)} / \sqrt{2}$ & \hspace*{0.05cm} $\omega_{\rm L0}^{(1)} - \omega_2 \pm \sqrt{2} g$ \hspace*{0.05cm} \\
		                   & $\ket{\lambda_1,\pm}$ & $\Omega_0^{(2)} / \sqrt{2}$ & \hspace*{0.05cm} $\omega_{\rm L0}^{(2)} - \omega_2 \pm \sqrt{2} g$ \hspace*{0.05cm} \\
		                   & $\ket{\lambda_0,\pm}$ & $\Omega_1$ & \hspace*{0.05cm} $\omega_{\rm L1} + \omega_1 - \omega_2 \mp g$ \hspace*{0.05cm} \\ \hline
		$\ket{-,0}$ & $\ket{\mu_1}$ & $\Omega_0^{(1)}$ & \hspace*{0.05cm} $\omega_{\rm L0}^{(1)} - \omega_2$ \hspace*{0.05cm} \\
		                 & $\ket{\mu_1}$ & $\Omega_0^{(2)}$ & \hspace*{0.05cm} $\omega_{\rm L0}^{(2)} - \omega_2$ \hspace*{0.05cm} \\
		                 & $\ket{\mu_0,\pm}$ & $\Omega_1$ & \hspace*{0.05cm} $\omega_{\rm L1} + \omega_1 - \omega_2 \mp g$ \hspace*{0.05cm} \\ \hline 
		$\ket{11,0}$ & $\ket{\lambda_1,\pm}$ & $\Omega_1$ & \hspace*{0.05cm} $\omega_{\rm L1} + \omega_1 - \omega_2 \mp \sqrt{2} g$ \hspace*{0.05cm} \\ \hline
	\end{tabular}
	\caption{Most relevant laser-driven transitions of the atom-cavity system in the dressed state picture. The table shows the respective ground and excited states and indicates the corresponding laser parameters.} \label{t_off_res2}
\end{table}

To make it nevertheless easy to identify the relevant laser Rabi frequencies and detunings, we now transform the laser Hamiltonian $\tilde H_{\rm L} (t)$ for the subspace ${\cal H}_0 \oplus {\cal H}_1$ into the interaction picture with respect to  the system Hamiltonian in Eq.~(\ref{Hsys}). Taking into account the eigenvalues of this Hamiltonian which can be found in Tables~\ref{g_states} and \ref{e_states} we obtain another time-dependent Hamiltonian from which we can directly read off the information which is relevant for the construction of an entangling scheme via cooling. The result of this calculation is summarised in Table \ref{t_off_res2} which shows all laser-driven transitions and states the corresponding relevant laser parameters. 

\subsection{Effect of spontaneous emission}

As has been illustrated already in Section~\ref{toy_model}, dissipation is an essential component of state preparation via cooling. In the atom-cavity system analysed in this section, dissipation can occur via the photon emission from the excited atomic state $|2 \rangle$ with the spontaneous decay rate $\Gamma$ and via the leakage of a photon through the cavity mirrors with the spontaneous decay rate $\kappa$. The conditional Hamiltonian that describes the time evolution of the atom-cavity system between photon emissions equals 
\begin{eqnarray} \label{Hcond}
	H_{\rm cond} &=& H_{\rm sys} + H_{\rm L}(t) - {{\rm i} \over 2} \hbar \Gamma \sum_{i=1}^{2} \vert 2 \rangle_{ii} \langle 2 \vert - {{\rm i} \over 2} \hbar \kappa \, c^{\dagger}c \, . \nonumber \\  
\end{eqnarray}
The first two terms in this equation are the system Hamiltonian $H_{\rm sys}$ in Eq.~(\ref{Hsys}) and the laser Hamiltonian $H_{\rm L}(t)$ in Eq.~(\ref{HL}). In case of an emission, the density matrix of the atom-cavity system changes up to normalisation into
\begin{eqnarray} \label{RR}
	\mathcal{R}(\varrho) &=& \sum_{j=0,1}\sum_{i=1,2} \Gamma_{j} \, \vert j \rangle_{ii} \langle2\vert \varrho \vert2\rangle_{ii}\langle j \vert + \kappa c\varrho c^{\dagger} \, , ~~
\end{eqnarray}
where $\Gamma_j$ denotes the spontaneous decay rate of the atomic 2--$j$ transition. The overall decay rate of the excited atomic state is given by $\Gamma = \Gamma_0 + \Gamma_1$. Overall, the time evolution of the system in the presence of spontaneous emission is described by master equations which are of exactly the same form as the master equations in Eq.~(\ref{t_master}).

\subsection{Appropriate laser parameters} \label{main2}

\begin{table}[t]
	\begin{tabular}{| c | c | c | c |}
		\hline
		~~ Ground ~~ & ~~ Excited ~~ & Rabi & ~~ Effective ~~ \\[-0.15cm]
		state & state  & ~~ frequency ~~ & ~~ detuning ~~ \\ \hline \hline
		$\ket{00,0}$ & $\ket{\lambda_0,+}$ & $\Omega_0^{(1)} / \sqrt{2}$ & $-2g$ \\
		                    & $\ket{\lambda_0,-}$ & $\Omega_0^{(1)} / \sqrt{2}$ & $0$ \\
		                    & $\ket{\lambda_0,\pm}$ & $\Omega_0^{(2)} / \sqrt{2}$ & $\mp g$ \\ \hline
		$\ket{+,0}$ & $\ket{\lambda_1,\pm }$ & $\Omega_0^{(1)} / \sqrt{2}$ & ~~ $\pm (\sqrt{2} \pm 1)g$ ~~ \\
		                  & $\ket{\lambda_1,\pm}$ & $\Omega_0^{(2)} / \sqrt{2}$ & $\mp \sqrt{2}g$ \\
		                  & $\ket{\lambda_0,\pm}$ & $\Omega_1$ & $-(\sqrt{2} \pm 1)g$ \\ \hline
		$\ket{-,0}$ & $\ket{\mu_1}$ & $\Omega_0^{(1)}$ & $-g$ \\
		                 & $\ket{\mu_1}$ & $\Omega_0^{(2)}$ & $0$ \\
		                 & $\ket{\mu_0,\pm}$ & $\Omega_1$ & $-(\sqrt{2} \pm 1)g$ \\ \hline
		$\ket{11,0}$ & $\ket{\lambda_1,+}$ & $\Omega_1$ & $-2\sqrt{2}g$ \\
		                    & $\ket{\lambda_1,-}$ & $\Omega_1$ & 0 \\
		\hline
	\end{tabular}
	\caption{Transitions between dressed states driven near resonance by the application of three lasers with Rabi frequency $\Omega_0^{(1)}$, $\Omega_0^{(2)}$ and $\Omega_1$.} \label{t_off_res}
\end{table}

As already mentioned above, the target state of the state preparation which we propose here is the maximally entangled atomic state $|+ \rangle$ in Eq.~(\ref{target}). In order to assure that this state becomes the stationary state of the atom-cavity system in Fig.~\ref{scheme}, we need to choose the laser frequencies $\omega_{\rm L0}^{(1)}$, $\omega_{\rm L0}^{(2)}$, and $\omega_{\rm L1}$ such that the $|+,0 \rangle$ experiences only off-resonant driving, while the states $|00,0 \rangle$, $|-,0 \rangle$, and $|11,0 \rangle$ couple resonantly to at least one of the three driving lasers. Having a closer look at Table \ref{t_off_res2}, we see that this applies, if we choose
\begin{eqnarray} \label{main}
	\omega_{\rm L0}^{(1)} &=& \omega_2 - g \, , \nonumber \\
	\omega_{\rm L0}^{(2)} &=& \omega_2 \, , \nonumber \\
	\omega_{\rm L1} &=& \omega_2 - \omega_1 - \sqrt{2} g \, .
\end{eqnarray}
Table \ref{t_off_res} shows the effect of this choice of laser frequencies on the sixteen transitions which need to be taken into account when designing the state preparation scheme proposed in this paper. 

\begin{figure}[t]
\begin{minipage}{\columnwidth}
\begin{center}
\resizebox{\columnwidth}{!}{\rotatebox{0}{\includegraphics{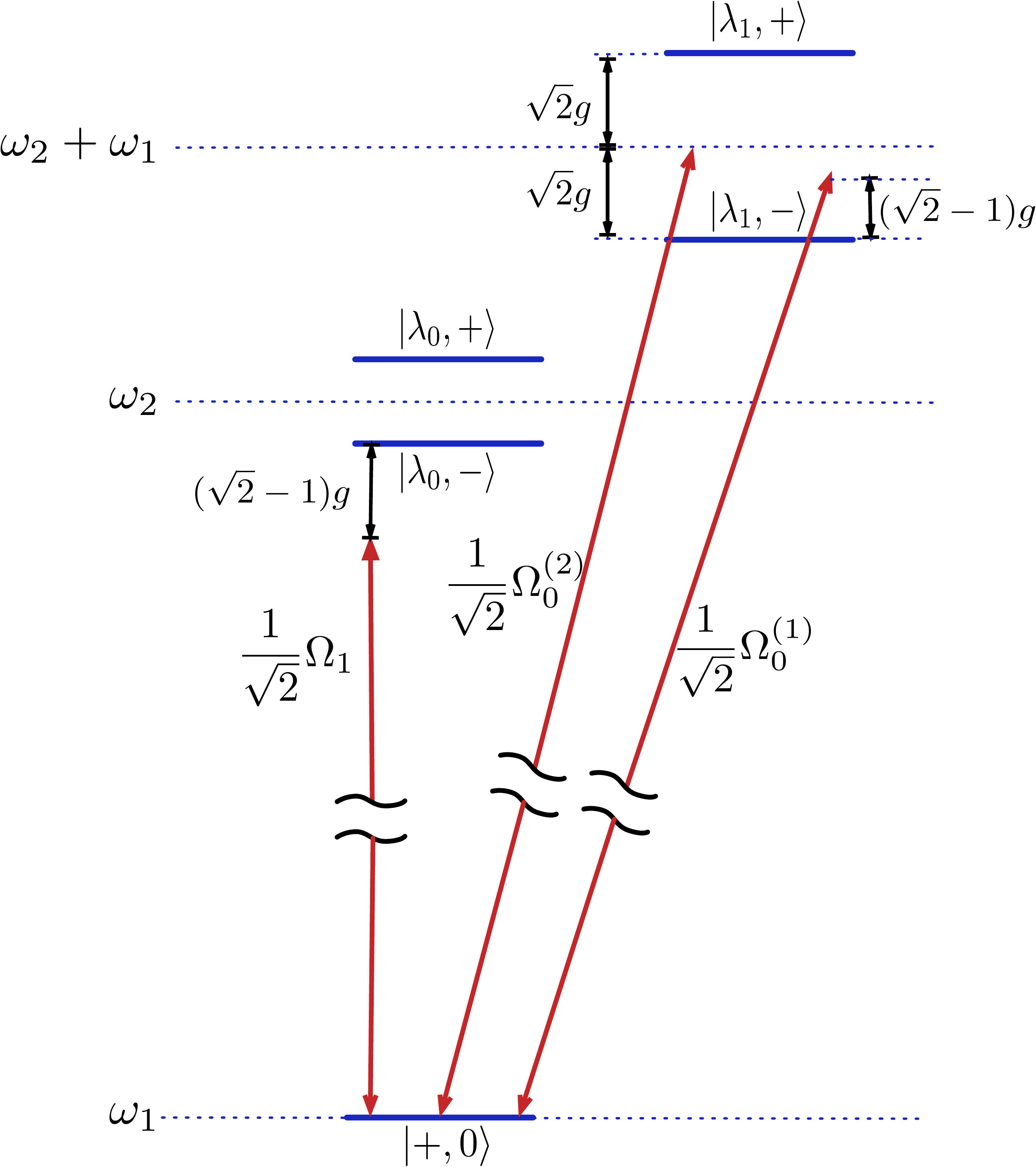}}}
\end{center}
\vspace*{-0.5cm} \caption{Level configuration showing the laser driving, Rabi frequencies, and detunings experienced by the target state $\ket{+,0}$ in the dressed state picture. For simplicity, we show only the least detuned couplings.} \label{off_res}
\end{minipage}
\end{figure}

The system Hamiltonian $H_{\rm sys}$ treats both atoms in exactly the same way. Its eigenvectors are therefore either symmetric or antisymmetric with respect to an exchange of the two atoms. The same applies to the effective laser Hamiltonian $\tilde H_{\rm L}(t)$. Since both atoms experience exactly the same Rabi frequencies, the lasers excite either transitions between two symmetric states or two anti-symmetric states. This allows us to consider the symmetric and the antisymmetric state space separately when analysing the effect of the laser driving in the dressed state picture of the atom-cavity system. There are three symmetric ground states and one antisymmetric ground state. These are $\{ \ket{00,0}, \, \ket{+,0}, \, \ket{11,0} \}$ and $\{ \ket{-,0} \}$ respectively.

\noindent \begin{figure*}[t]
\begin{minipage}{2\columnwidth}
\begin{center}
\resizebox{\columnwidth}{!}{\rotatebox{0}{\includegraphics{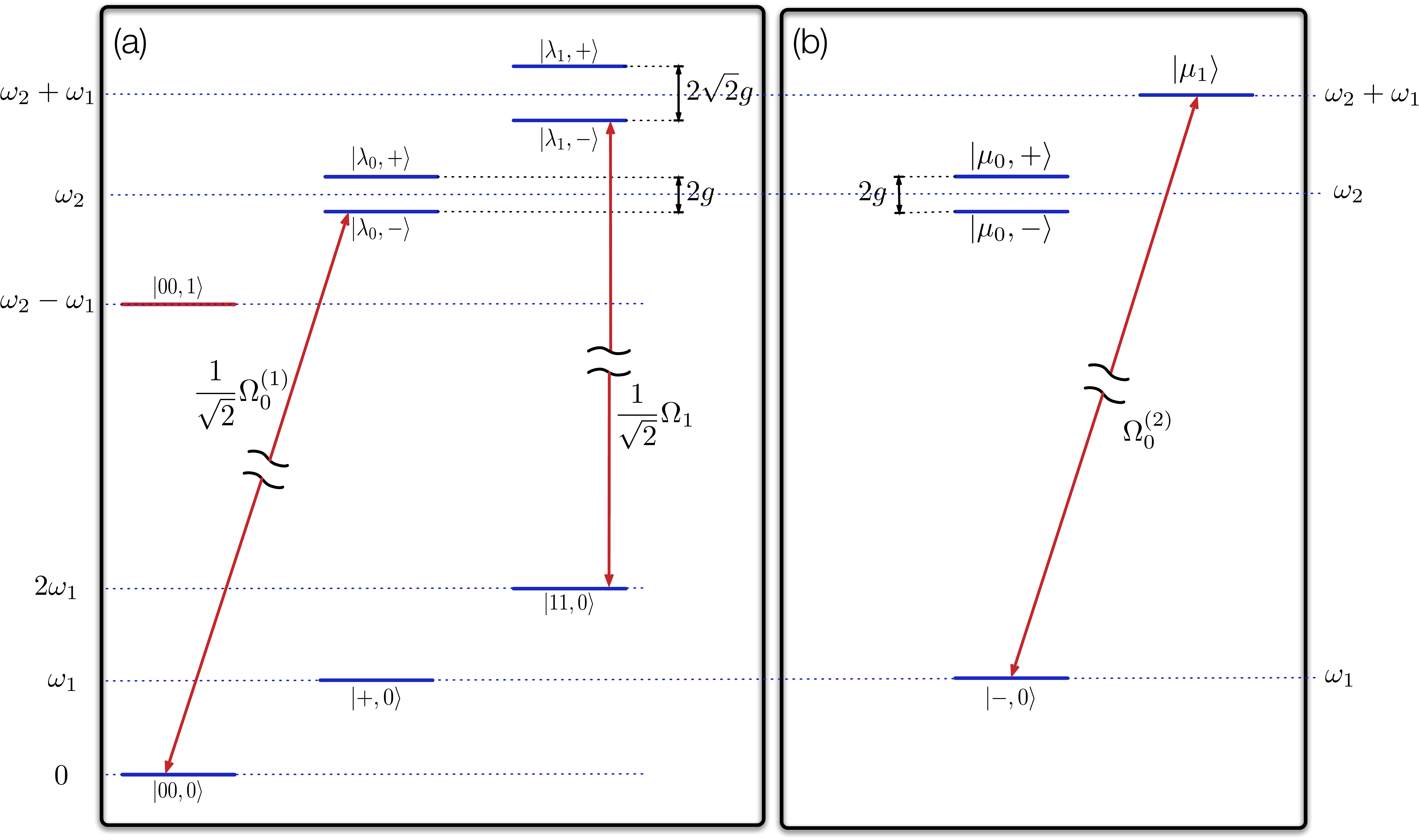}}}
\end{center}
\caption{Level configuration showing all the resonantly driven transitions in the dressed state picture, their Rabi frequencies, and their detunings in the subspace with zero or one excitation in $|2 \rangle$ or the cavity mode. (a) Symmetric subspace with ground states $\ket{00,0}$, $\ket{+,0}$, and $\ket{11,0}$. (b) Antisymmetric subspace with ground state $\ket{-,0}$.} \label{sym}
\end{minipage}
\end{figure*}

Fig.~\ref{off_res} illustrates the laser driving experienced by the target state $|+,0 \rangle$. Since this state is a symmetric state, the relevant level configuration involves only the target state and the four symmetric states with one excitation in $|2 \rangle$ or the cavity mode. As one can see from Table \ref{t_off_res}, in the dressed state picture, these three lasers involve $|+, 0 \rangle$ in six different transitions. For simplicity, we show only the least detuned couplings for each laser. We see that the target state $|+,0 \rangle$ experiences indeed only off-resonant driving. The smallest and therefore most relevant detuning is given by  
\begin{eqnarray} \label{delta_min}
\delta_{\rm min} &=& (\sqrt{2} - 1) g \, ,
\end{eqnarray}
as long as the frequency $\omega_1$ is sufficiently larger than the atom-cavity coupling constant $g$. All other states with no excitation are resonantly driven by one laser field. This is illustrated in Fig.~\ref{sym} which shows the resonant transitions in the symmetric and the antisymmetric subspace separately. One laser couples $\ket{00,0}$ to $\ket{\lambda_0,-}$ with Rabi frequency $\Omega_0^{(1)}/\sqrt{2}$. Another laser couples $\ket{-,0}$ to $\ket{\mu_1}$ with Rabi frequency $\Omega_0^{(2)}/\sqrt{2}$, while a third laser drives $\ket{11,0}$ into $\ket{\lambda_1,-}$ with Rabi frequency $\Omega_1$. In principle, we would like these lasers which empty unwanted states to be relatively strong. However, it is not possible to increase them without increasing also the Rabi frequencies for the off-resonant driving of the target state shown in Fig.~\ref{off_res}. 

\subsection{Fidelities and cooling rates for constant laser driving} \label{main3}

Eq.~(\ref{delta_min}) shows that the minimum detuning experienced by the target state $\delta_{\rm min}$ depends only on the atom-cavity coupling constant $g$. Eq.~(\ref{toy_regime}) in Section \ref{toy_model} therefore suggests that the stationary state of the atom-cavity system in Fig.~\ref{scheme} is to a very good approximation given by the state $|+,0 \rangle$ as long as
\begin{eqnarray} \label{condi}
\omega_1 \, \gg \, g \, \gg \, \Gamma, \, \kappa, \, \Omega_0^{(1)}, \, \Omega_0^{(2)}, \, \Omega_1 \, .
\end{eqnarray}
In other words, for this parameter regime we can expect the atoms to be with a very high fidelity in the maximally entangled state $|+ \rangle$ in Eq.~(\ref{target}) after a certain transition time $t$. A comparison with the toy-model state preparation scheme in Section \ref{toy_model} even yields approximate solutions for the fidelity ${\rm F}$ and the cooling rate $\gamma_{\rm c}$ of the proposed entangling scheme. 

For simplicity, we assume in the following that all three cooling lasers have the same Rabi frequency $\Omega $ and that the two atomic decay rates, $\Gamma_0$ and $\Gamma_1$, are equal. The only remaining spontaneous decay rates are the spontaneous atom decay rate $\Gamma$ and the cavity photon leakage rate $\kappa $. For example, the symmetric state $(\ket{\lambda_0,+} + \ket{\lambda_0,-})/\sqrt{2}$ with one atom in $| 2 \rangle$ has the spontaneous decay rate $\Gamma$, whilst the state $(\ket{\lambda_0,+} - \ket{\lambda_0,-})/\sqrt{2}$ with one photon in the cavity decays with $\kappa$. We infer from this that the fidelity of the proposed entangling scheme depends on the size of both decay rates. Taking this into account, we replace the spontaneous decay rate $\Gamma$ in Eq.~(\ref{gamma_c2}) in the following with the average of $\kappa$ and $\Gamma$, 
\begin{eqnarray}
\Gamma &\longrightarrow & {1 \over 2} (\kappa + \Gamma) \, .
\end{eqnarray}
The analog of the laser detuning $\Delta$ in the toy-model state preparation scheme is the detuning $\delta_{\rm min}$ in Eq.~(\ref{delta_min}) which is the minimum laser detuning experienced by the target state $|+,0 \rangle$ during the cooling process. Taking into account that
\begin{eqnarray}
\Delta &\longrightarrow & \delta_{\rm min} \, ,
\end{eqnarray}
Eqs.~(\ref{gamma_c2}) and (\ref{t_stationary2}) suggest that $\gamma_{\rm c}$ and ${\rm F}$ are to a very good approximation given by
\begin{eqnarray} \label{gamma_c3}
	\gamma_{\rm c} &=& \frac{2 \Omega^2 (\kappa + \Gamma)}{12\Omega^2+(\kappa+\Gamma)^2} \, , \nonumber \\
	{\rm F} &=& 1 - \frac{12 \Omega^2 + (\kappa + \Gamma)^2}{16 (\sqrt{2}-1)^2 g^2} \, .
\end{eqnarray}
This result confirms Eq.~(\ref{condi}) which suggests that fidelities ${\rm F}$ close to one are only obtained when $g$ is much larger than all other system parameters. 

\begin{figure}[t]
\begin{minipage}{\columnwidth}
\begin{center}
\resizebox{\columnwidth}{!}{\rotatebox{0}{\includegraphics{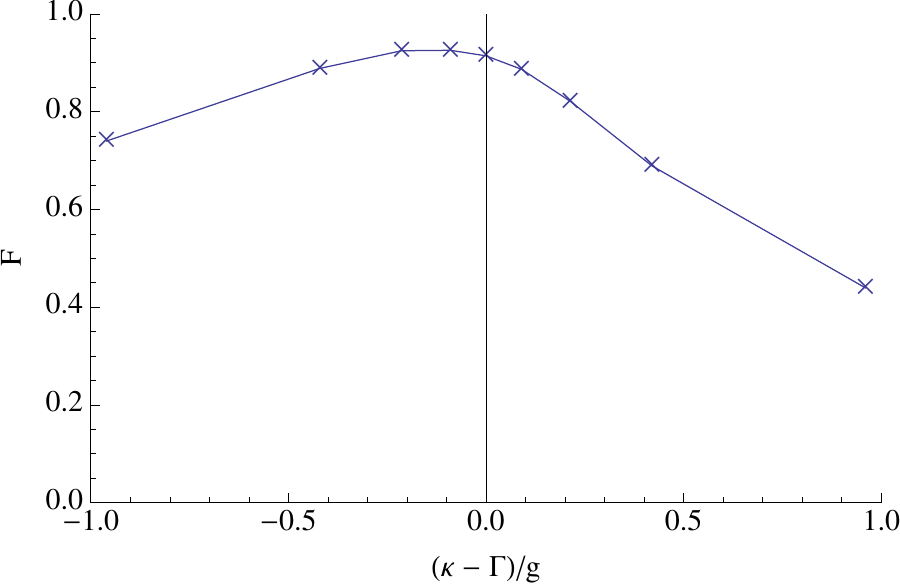}}}
\end{center}
\vspace*{-0.5cm} \caption{Stationary state fidelity ${\rm F}$ of the proposed entangling scheme for different spontaneous decay rates $\kappa$ and $\Gamma$ and $\Omega = 0.03 \, g$ obtained from a quantum jump simulation. Here $\kappa$ and $\Gamma$ are varied such that the cooperativity parameter $C$ remains constant at $C=25$.} \label{Kvar}
\end{minipage}
\end{figure}

The remainder of this paper confirms these approximate solutions with the help of a numerical analysis of the proposed entangling scheme. The following analysis is based on the quantum jump approach \cite{Hegerfeldt} which allows us to simulate all the possible trajectories of the atom-cavity system in Fig.~\ref{scheme}. By averaging over many trajectories, we obtain an approximate solution of the master equation in Eq.~(\ref{t_master}). To calculate the no-photon time evolution of the system we use the conditional Hamiltonian $H_{\rm cond}$ in Eq.~(\ref{Hcond}). In case of a photon emission we reset the atom-cavity system such that its state after a photon emission is on average given by $\mathcal{R}(\varrho) \Delta t$ with $\mathcal{R}(\varrho)$ as in Eq.~(\ref{RR}). For simplicity, we consider only relatively small Rabi frequencies. In this way we avoid the population of highly excited states. The population of such states is not expected to decrease the fidelity of the final state since they decay relatively rapidly. However, in this way we can restrict the size of the Hilbert space which has to be taken into account during simulations to states with at most three photons in the cavity.

\begin{figure}[t]
\begin{minipage}{\columnwidth}
\begin{center}
\resizebox{\columnwidth}{!}{\rotatebox{0}{\includegraphics{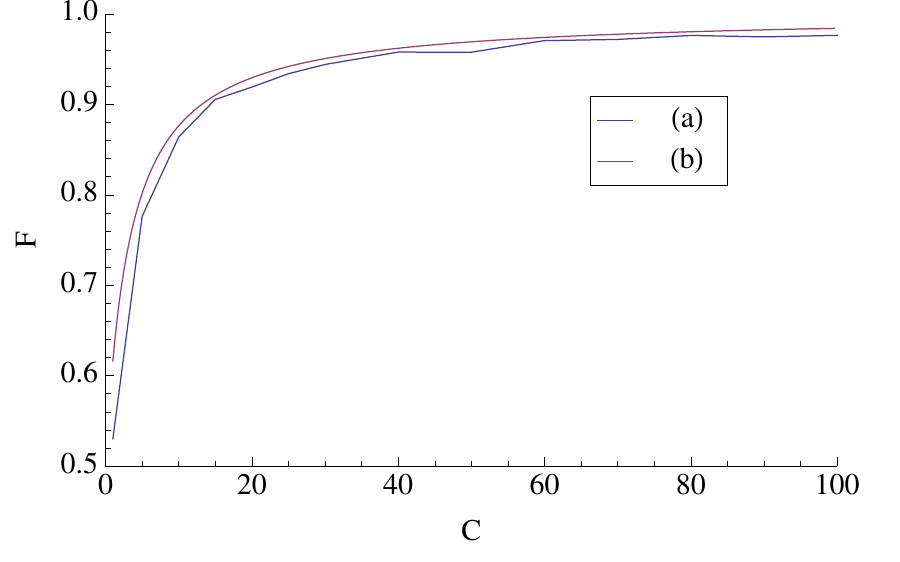}}}
\end{center}
\vspace*{-0.5cm} \caption{Stationary state fidelity ${\rm F}$ of the proposed entangling scheme as a function of the cooperativity parameter $C$ for $\Omega = 0.03 \, g$ and $\kappa = 2 \Gamma$. (a) Numerical solution of the time evolution of the system using the quantum jump approach. (b) Analytical result in Eq.~(\ref{gamma_c3}).} \label{fidelity}
\end{minipage}
\end{figure}

Let us first have a closer look at how changing the relative size of $\kappa$ with respect to $\Gamma$ affects the fidelity ${\rm F}$ of the state preparation. Fig.~\ref{Kvar} shows ${\rm F}$ for different spontaneous decay rates $\kappa$ and $\Gamma$. These are chosen such that the single-atom cooperativity parameter $C$ in Eq.~(\ref{C}) remains constant at $C = 25$. Each data point represents the average fidelity calculated from a time series like the one shown in Fig.~\ref{C25_compare}. As a result we find that the proposed state preparation scheme works best when $\Gamma - \kappa = 0.15 \, g$. This implies that ideally one should have
\begin{eqnarray}\label{O(t)4}
	\kappa & = &2 \Gamma \, ,
\end{eqnarray}
when $C = 25$. We therefore assume that this applies in the remainder of this section.

The main result of this subsection is an estimation of the stationary state fidelity of the maximally entangled atomic state $|+ \rangle$ which can be achieved with the proposed entangling scheme. To establish this numerically, we use a series of time evolutions such as those shown in Fig.~\ref{C25_compare} and average over the fidelity once the system is approximately in its stationary state. Fig.~\ref{fidelity} shows the stationary state fidelity of the target state as a function of the cooperativity parameter $C$ for a constant laser Rabi frequency $\Omega $. The numerical results are compared with the analytical result for ${\rm F}$ in Eq.~(\ref{gamma_c3}). Indeed we find very good agreement between analytical and numerical results. It is clear from Fig.~\ref{fidelity} that the achievable fidelity ${\rm F}$ increases rapidly with increasing cooperativity parameter $C$. However, fidelities above $90\%$ are possible, even for a cooperativity parameter $C$ as low as 20.

\subsection{Minimising the state preparation time} \label{main10}

\begin{figure}[t]
\begin{minipage}{\columnwidth}
\begin{center}
\resizebox{\columnwidth}{!}{\rotatebox{0}{\includegraphics{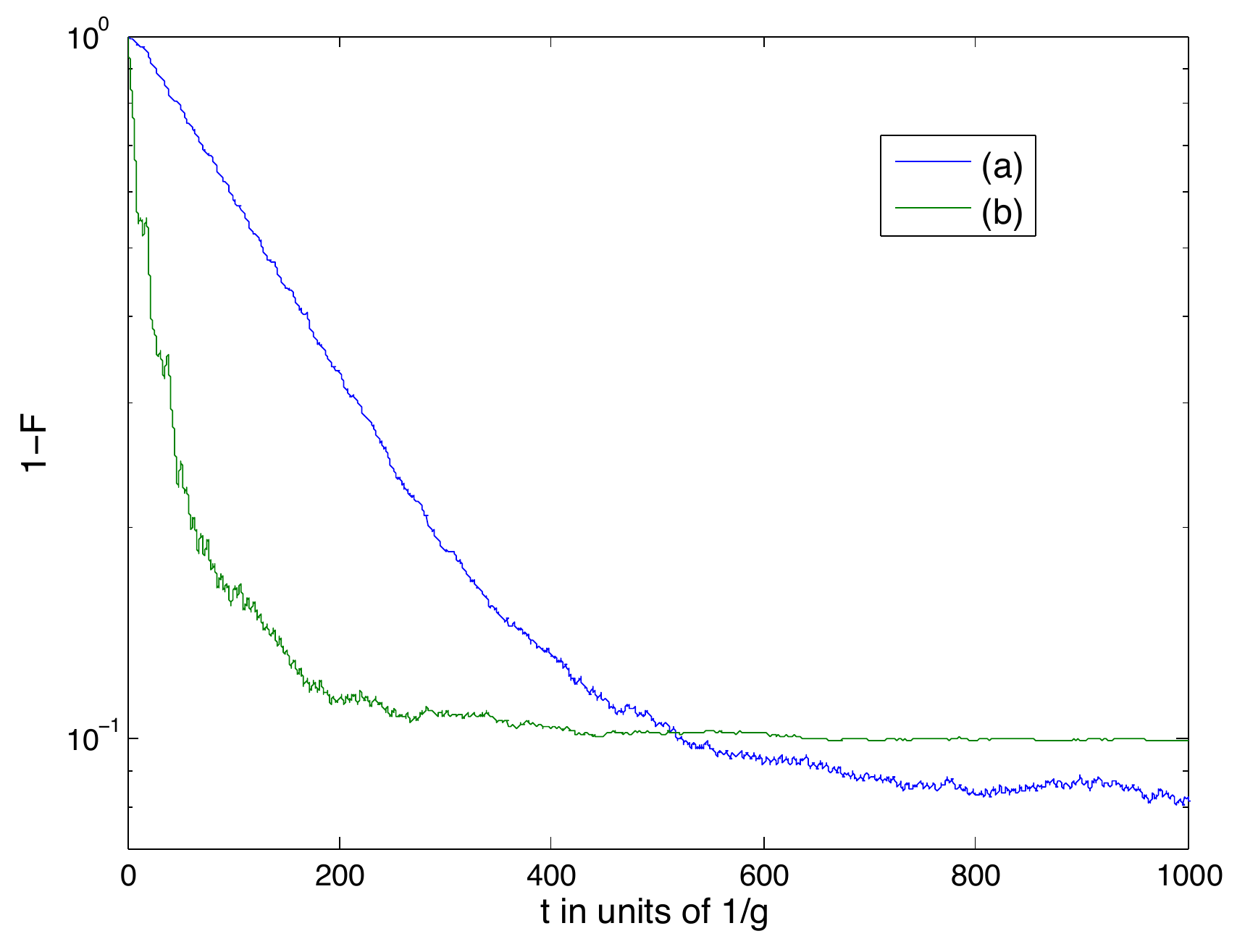}}}
\end{center}
\vspace*{-0.5cm} \caption{(a) Logarithmic plot of the time dependence of the distance $1-{\rm F}$ from the target state $|0 \rangle$ for the case where the system is initially in $\ket{00,0}$ and where $\kappa = 2 \Gamma = 0.2 \, g$ and $\Omega = 0.03 \, g$. (b) Same as (a) but for a time dependent laser pulse with a Rabi frequency $\Omega(t)$ as in Eq.~(\ref{O(t)2}) with $\Omega_0 = 0.015 \, g$.} \label{C25_compare}
\end{minipage}
\end{figure}

As in Section \ref{toy_model}, we find that a relatively large cooling rate $\gamma_{\rm c}$ requires relatively large Rabi frequencies. At the same time, we only obtain a fidelity ${\rm F}$ close to unity for very small Rabi frequencies. In order to maximise the fidelity of the state preparation while maintaining a substantial cooling rate, we therefore proceed in the following as in Section \ref{choosing} and assume a time dependent Rabi frequency $\Omega$. Similarly as in Eq.~(\ref{toy_Ot}), we assume in the following that
\begin{eqnarray} \label{O(t)2}
	\Omega(t) &=& {6 \Omega_0 \over \left( 1 + \gamma_{\rm c}(0) t \right)^{2}} \, ,
\end{eqnarray}
where $\gamma_{\rm c}(0)$ denotes the cooling rate of the entangling scheme for the initial Rabi frequency $\Omega(0) =\Omega_0$. Fig.~\ref{C25_compare} confirms that choosing a time-dependent Rabi frequency $\Omega$ indeed improves the speed of the entangling scheme without sacrificing much of its quality.

\section{Conclusions} \label{conc}

In this paper, we propose an entangling scheme for two atoms trapped inside an optical cavity. Each atom should contain a $\Lambda$-like level configuration with the ground states $|0 \rangle$ and $|1 \rangle$ forming one qubit and an excited state $|2 \rangle$ (cf.~Fig.~\ref{scheme}). Three laser fields should be applied simultaneously. Two of them continuously drive the 0--2 transition which is in resonance with the cavity mode, while the third laser drives the 1--2 transition. Individual laser addressing of the atoms is not required. Most importantly, the laser detunings should be chosen as proposed in Eq.~(\ref{main}) in Section \ref{main2}. As a result, the maximally entangled atomic ground state $|+,0 \rangle$ with no photons in the cavity becomes the stationary state of the atom-cavity system. To complete the state preparation, the laser fields should be turned off after a certain transition time. The presence of non-zero spontaneous decay channels, i.e.~the leakage of photons through the cavity mirrors and direct spontaneous emission from the atoms, are essential for the scheme to work.

The proposed state preparation scheme is a concrete realisation of a recent proposal to cool atoms into entangled state \cite{Vacanti:2009p5419}. Choosing the laser detunings as proposed in Eq.~(\ref{main}) guarantees that only the target state $|+,0 \rangle$ experiences off-resonant driving. All other states with no population in $|2 \rangle$ and in the cavity mode interact resonantly with one of the three applied laser fields. As in laser sideband cooling \cite{cool}, this makes it much more likely for the atom-cavity system to decay into the target state than being driven out of it. As a result, most of the population of the system accumulates in $|+,0 \rangle$ with both atoms in a well-defined, maximally entangled state.
Since the relevant detuning of this state is essentially given by the atom-cavity coupling constant $g$, the analogy to laser sideband cooling suggests that the scheme works best when all other system parameters are much smaller than $g$. 

Due to laser driving with three different laser fields, it is not possible to solve the time evolution of the proposed entangling scheme analytically. The reason is that there is no interaction picture in which the Hamiltonian of the system becomes time independent. To obtain at least approximate analytical solutions for the cooling rate $\gamma_{\rm c}$ and the stationary state fidelity ${\rm F}$ (cf.~Eq.~(\ref{gamma_c3})), Section \ref{toy_model} discusses a closely related state preparation scheme for a much simpler analytically tractable toy-model. A comparison with this toy-model provides much insight into the state preparation via cooling as well as analytical results. These are confirmed in Section \ref{main3} by extensive numerical solutions of the time evolution of the atom-cavity system in Fig.~\ref{scheme}. Section \ref{main10} finally suggests a method to speed up the state preparation without sacrificing its fidelity by using time-dependent laser fields with rapidly decreasing Rabi frequencies (cf.~Eq.~(\ref{O(t)2})).

Compared to other recent entangling schemes for atom-cavity systems \cite{Sorensen,Wang2}, the scheme proposed here predicts higher fidelities for the same experimental parameters. As illustrated in Fig.~\ref{fidelity} in Section \ref{main3}, it can achieve fidelities above $90\%$ even when the single-atom cooperativity parameter $C$ is as low as 20.
With a cooperativity parameter $C = 25$ we can achieve a fidelity of $93\%$, while Ref.~\cite{Sorensen} predicts fidelities above $92\%$ only for $C > 50$. Ref.~\cite{Sorensen} uses a similar level scheme as our proposal but with the addition of a driven microwave transition between the triplet states. Ref.~\cite{Wang2} requires the presence of a magnetic field gradient to produce the required level splittings to cool atoms into an entangled state. Compared to other quantum computing schemes {\em using} dissipation \cite{Pellizzari,zoller2,Beige,Pachos,Rempe,zoller3,sean,Lim,Metz}, the state preparation scheme discussed here no longer relies on the detection of single photons or macroscopic fluorescence signals to herald the success of the state preparation. Its implementation might already be in reach with current technology. \\[0.5cm]
{\em Acknowledgement.} J. B. acknowledges financial support from the European Commission of the European Union under the FP7 STREP Project HIP (Hybrid Information Processing). A. B. acknowledges a James Ellis University Research Fellowship from the Royal Society and the GCHQ. S. S. I. and B. T. T. have been supported by the European Union Research and Training Network EMALI.


\begin{thebibliography}{99}
\bibitem{Rempe0}
J. Bochmann, M. Muecke, C. Guhl, S. Ritter, G. Rempe, and D. L. Moehring, Phy. Rev. Lett. {\bf 104}, 203601 (2010).

\bibitem{Walther}
M. Keller, B. Lange, K. Hayasaka, W. Lange, and H. Walther, J. Mod. Opt. {\bf54}, 1607 (2007).

\bibitem{Chapman}
K. M. Fortier, Y. Kim, M. J. Gibbons, P. Ahmadi, and M. S. Chapman, Phys. Rev. Lett. {\bf 98}, 233601 (2007).

\bibitem{Meschede}
M. Khudaverdyan, W. Alt, T. Kampschulte, S. Reick, A. Thobe, A. Widera, and D. Meschede, Phys. Rev. Lett. {\bf 103}, 123006 (2009). 

\bibitem{Hinds}
M. Trupke, J. Goldwin, B. Darquié, G. Dutier, S. Eriksson, J. Ashmore, and E. A. Hinds, Phys. Rev. Lett. {\bf 99}, 063601 (2007).

\bibitem{Reichel}
R. Gehr, J. Volz, G. Dubois, T. Steinmetz, Y. Colombe, B. L. Lev, R. Long, J. Esteve, and J. Reichel, Phys. Rev. Lett. {\bf 104}, 203602 (2010).

\bibitem{Pellizzari}
T. Pellizzari, S. A. Gardiner, J. I. Cirac, and P. Zoller, Phys. Rev. Lett. {\bf 75}, 3788 (1995).

\bibitem{zoller2}
M. B. Plenio, S. F. Huelga, A. Beige, and P. L. Knight, Phys. Rev. A {\bf 59}, 2468 (1999).

\bibitem{Beige}
A. Beige, D. Braun, B. Tregenna, and P. L. Knight, Phys. Rev. Lett. {\bf 85}, 1762 (2000).

\bibitem{zheng}
S.-B. Zheng and G. C. Guo, Phys. Rev. Lett. {\bf 85}, 2392 (2000).

\bibitem{Pachos}
J. Pachos and H. Walther, Phys. Rev. Lett. {\bf 89}, 187903 (2002).

\bibitem{you}
X. X. Yi, X. H. Su, and L. You, Phys. Rev. Lett. {\bf 90}, 097902 (2003).

\bibitem{Rempe}
C. Marr, A. Beige, and G. Rempe, Phys. Rev. A {\bf 68}, 033817 (2003).

\bibitem{zoller3}
D. E. Browne, M. B. Plenio, and S. F. Huelga, Phys. Rev. Lett. {\bf 91}, 067901 (2003). 

\bibitem{Lim}
Y. L. Lim, A. Beige, and L. C. Kwek, Phys. Rev. Lett. {\bf 95}, 030505 (2005).

\bibitem{sean}
S. D. Barrett and P. Kok, Phys. Rev. A {\bf 71}, 060310(R) (2005).

\bibitem{Metz}
J. Metz, M. Trupke, and A. Beige, Phys. Rev. Lett. \textbf{97}, 040503 (2006).

\bibitem{Milburn}
S. Schneider and G. J. Milburn, Phys. Rev. A {\bf 65}, 042107 (2002).

\bibitem{Kraus-Cirac}
B. Kraus and J. I. Cirac, Phys. Rev. Lett. {\bf 92}, 013602 (2004).

\bibitem{Diehl:2008p7796} 
S. Diehl, A. Micheli, A. Kantian, B. Kraus, H. P. B\"uchler, and P. Zoller, Nature Phys. {\bf 11} 878 (2008).

\bibitem{Kraus:2008p5470} 
B. Kraus, H. P. B\"uchler, S. Diehl, A. Kantian, A. Micheli, and P. Zoller, Phys. Rev. A {\bf 78}, 042307 (2008).

\bibitem{Verstraete:2009p3815} 
F. Verstraete, M. M. Wolf, and J. I. Cirac, Nature Phys. {\bf 9}, 633 (2009).

\bibitem{Vacanti:2009p5419} 
G. Vacanti and A. Beige, New. J. Phys. {\bf 11}, 083008 (2009).

\bibitem{Wang}
X. T. Wang and S. G. Schirmer, Phys. Rev. A {\bf 80}, 042305 (2009).

\bibitem{Viola}
F. Ticozzi and L. Viola, Automatica {\bf 45}, 2002 (2009). 

\bibitem{Cho}
J. Cho, S. Bose, and M. S. Kim, Phys. Rev. Lett. {\bf 106}, 020504 (2011).

\bibitem{Sorensen}
M. J. Kastoryano, F. Reiter, and A. S. S{\/o}rensen, Phys. Rev. Lett. {\bf 106}, 090502 (2011).

\bibitem{Wang2}
X. T. Wang and S. G. Schirmer, {\em Generating maximal entanglement between non-interacting atoms by collective decay and symmetry breaking}, arXiv:1005.2114 (2010).

\bibitem{cool}
D. J. Wineland and W. M. Itano, Phys. Rev. A {\bf 20}, 1521 (1979).  

\bibitem{Hegerfeldt}
G. C. Hegerfeldt, Phys. Rev. A {\bf 47}, 449 (1993).
\end{thebibliography}
\end{document}